\documentstyle[aps,prd]{revtex}
 \def\lsim{\raise0.3ex\hbox{$\;<$\kern-0.75em\raise-1.1ex
 \hbox{$\sim\;$}}}
 \def\gsim{\raise0.3ex\hbox{$\;>$\kern-0.75em\raise-1.1ex
 \hbox{$\sim\;$}}}

\begin{document}

\title{Probing supernova physics with neutrino oscillations}
\author{
Hisakazu Minakata$^1$\thanks{minakata@phys.metro-u.ac.jp}, 
Hiroshi Nunokawa$^2$\thanks{nunokawa@ift.unesp.br}, 
Ricard Tom{\`a}s$^{3, 4}$\thanks{ricard@mppmu.mpg.de}
and J.~W.~F. Valle$^4$\thanks{valle@ific.uv.es}
}
\address{
$^1$Department of Physics, Tokyo Metropolitan University,
1-1 Minami-Osawa, Hachioji, Tokyo 192-0397, Japan
} 
\address{
$^2$Instituto de F\'{\i}sica Te{\'o}rica, Universidade Estadual Paulista, 
Rua Pamplona 145 \\ 
01405-900 S{\~a}o Paulo, SP Brazil
}
\address{
$^3$Max-Planck-Institut f\"ur Physik (Werner-Heisenberg-Institut), 
F\"ohringer Ring 6, 80805 M\"unchen, Germany
}
\address{
$^4$Institut de F\'{\i}sica Corpuscular -- C.S.I.C., 
  Universitat de Val{\`e}ncia, Edifici Instituts, \\ 
Aptdo.\ 22085, E--46071 Val{\`e}ncia, Spain
}

\date{July 10, 2002 }
\maketitle

\hfuzz=25pt

\begin{abstract}

We point out that solar neutrino oscillations with large mixing angle
as evidenced in current solar neutrino data have a strong impact on
strategies for diagnosing collapse-driven supernova (SN) through
neutrino observations.
Such oscillations induce a significant deformation of the energy
spectra of neutrinos, thereby allowing us to obtain otherwise
inaccessible features of SN neutrino spectra.
We demonstrate that one can determine temperatures and luminosities of
non-electron flavor neutrinos by observing $\bar{\nu}_{e}$ from
galactic SN in massive water Cherenkov detectors by the charged
current reactions on protons.

\end{abstract}

\vspace{0.5cm}

\section{Introduction}

The historical observation of neutrinos from the supernova (SN) 1987A
at the Kamiokande \cite{Kam} and IMB \cite{IMB} detectors had a great
impact and confirmed the basic picture of stellar collapse and SN
explosion~\cite{janka}.  However, the number of observed events was
too small to draw definite conclusions about the explosion mechanism
or detailed properties of SN neutrinos.
Hopefully, the next SN in our galaxy will be detected by several
massive neutrino detectors, currently operating or planned.  Such
observations of neutrinos from a galactic SN should provide us a far
more detailed information on properties of SN neutrinos, as well as
the newly formed neutron star, thus bringing substantial progress to
our understanding of stellar collapse.

The discovery of atmospheric neutrino oscillations at Super-Kamiokande
(SK)~\cite{SKatm} implies that neutrinos have masses and different
neutrino flavors mix, and it is supported by the results from the
first long-baseline neutrino oscillation experiment, K2K \cite{K2K}.
Moreover, a clear evidence for solar neutrino oscillations into mu/tau
neutrinos has been obtained by combining the Sudbury Neutrino
Observatory (SNO) charged current (CC) measurement \cite{SNO1} with
elastic scattering measurement at SK \cite{SK} and the most
recent {\it in situ} CC and neutral current (NC) measurements at SNO
\cite{SNO2}.

Neutrino oscillations add a new ``complication'' to the diagnostics of
SN neutrinos, since it is no longer true that a neutrino $\nu_\alpha$
leaving the neutrinosphere with definite flavor $\alpha$ will be
detected on Earth as the same neutrino species.
Then, the task of extracting information such as the original 
of neutrino spectra in the SN core from terrestrial observations, 
requires solving an ``inverse problem''.

In this letter, we point out that rather than a ``complication'' the
neutrino oscillation provides a new powerful tool for probing
otherwise inaccessible features of SN neutrino spectra.  This holds if
the mixing angle responsible for the solar neutrino oscillation is
large \cite{solaranalysis} as clearly indicated by the current solar
neutrino data.

We show that a high statistics observation of $\bar{\nu}_e$ through
the CC reaction $\bar{\nu}_e + p \to n + e^+$ will enable us to
extract not only the original temperature of $\bar{\nu}_e$ but also
that of $\bar{\nu}_{\mu(\tau)}$ at the neutrinosphere, as well as
their time integrated luminosities.
In order to determine these parameters we employ a $\chi^2$ method to
``separate'' two different neutrino spectra at the neutrinosphere with
different temperatures and luminosities. Hereafter we use a collective
notation $\nu_{x}$ ($\bar{\nu}_{x}$) for $\nu_{\mu(\tau)}$
($\bar{\nu}_{\mu(\tau)}$) because they cannot be distinguished by
their physical properties inside the SN.

We focus on $\bar{\nu}_e$ observation for the following reasons: (i)
this is expected to be the channel with highest statistics in a large
water Cerenkov detectors such as SK, Hyper-Kamiokande (HK) or UNO
detectors, which are under consideration~\cite{onemegaton}.  Neutrinos
from a galactic SN at 10 kpc would produce 7000-10000 events at SK and
2-3 $\times 10^5$ events at HK~\cite{totsuka}.  (ii) the large solar
neutrino mixing angle necessarily implies that the $\bar{\nu}_e$
spectrum observed at the Earth is a strong mixture of two originally
different SN neutrino spectra of $\bar{\nu}_e$ and $\bar{\nu}_{x}$.
This fact has been used to derive various constraints on neutrino
mixing parameters ~\cite{SSB94,MN00,Kachelriess:2001sg}.
Complementary information may be obtained also through direct
detection of $\bar{\nu}_{x}$ through the NC reaction either at SNO or
KamLAND~\cite{SNO/KLAND}.

In addition to generic features of SN neutrino spectra obtained in SN
simulations~\cite{janka,MayleWilson,SNsimu,JH89,Raffelt} we will take
into account some new features indicated by recent studies. 
Most importantly, they include a new parameter which characterize 
the departure from the equipartition of integrated luminosities to  
electron and other flavor neutrinos, the quantity of greatest 
uncertainty in SN simulations.  
(See below.)

\section{Supernova neutrinos; basic properties}

We now briefly summarize the features of neutrino spectra relevant 
for our work.
In a SN explosion driven by gravitational collapse, about 99 \% of the
total binding energy of the neutron star, $E_b\simeq 3 \times 10^{53}$
erg, is released in the form of neutrinos during the first $\sim 10$
seconds after the onset of core collapse.
It is well known that the time-dependent energy spectrum of each
neutrino species can be approximated by a ``pinched'' Fermi-Dirac (FD)
distribution \cite{JH89}.
In this work, we assume that the time-integrated spectra can also be
well approximated by the pinched FD distributions with an effective
degeneracy parameter $\eta$
\begin{equation}
f(E)
\propto
\frac{E^2}{e^{E/T-\eta}+1},
\label{eq:FD}
\end{equation}
where $E$ denotes the neutrino energy, and $T$ the effective
temperature. We have checked the validity of this approximation by
using results of 20 $M_\odot$ simulation by Mayle and Wilson
\cite{MayleWilson,TSDW}; we take in our analysis $\eta = 2.6$ and
$\eta = 0$ for $\bar{\nu}_e$ and $\bar{\nu}_{x}$, respectively, which
reproduce well the time-integrated spectra.

Because of the hierarchy in strength of interactions with the
surrounding matter neutrino temperatures obey $\langle T_{\nu_e}
\rangle < \langle T_{\bar{\nu}_e} \rangle < \langle T_{{\nu}_{x}}
\rangle \simeq \langle T_{\bar{\nu}_{x}} \rangle$ as confirmed by
various SN simulations.
Typical values of the average energies of the time-integrated neutrino
spectra obtained
are 
$\langle E_{{\nu}_e} \rangle \sim 12$ MeV, 
$\langle E_{\bar{\nu}_e} \rangle \sim 15$ MeV 
and $\langle E_{{\nu}_x} \rangle \simeq
\langle E_{\bar{\nu}_x} \rangle \sim 24$ MeV. 
We introduce, for later use, a new parameter $\tau_E$ which is defined
as the ratio of the average energies of the time-integrated neutrino
spectra, 
\begin{equation}
\tau_E \equiv \frac {\langle E_{\bar{\nu}_\mu}\rangle}
{\langle E_{\bar{\nu}_e}\rangle }.
\end{equation}
SN simulations indicate $\tau_E \sim 1.25-2.0$.

Recent studies and SN simulations indicate various new features which
were not taken into account in previous studies.
There are three effects at least.  The first is the possibility of a
gross violation of equality in integrated luminosities of 
$\nu_e/\bar{\nu}_e$ and $\nu_x/\bar{\nu}_{x}$ by up to $\sim 50$ \%
\cite{mezza}.  The second is a violation of equality of physical
properties of $\nu_{x}$ to $\bar{\nu}_{x}$ due to the effects of weak
magnetism~\cite{horo}.  The third is possible difference in integrated
luminosity between $\bar{\nu}_e$ and $\nu_e$ \cite{mezza,BJKRR,RJ00}.

The actual situation regarding violation of equipartition 
can be less dramatic since the $\nu_e \bar{\nu_e}$ annihilation 
process enhances the $\nu_{x}$/$\bar{\nu}_{x}$ luminosity, 
as recently shown by Buras {\it et al.}~\cite{BJKRR}. 
However, we prefer to introduce a free fit parameter $\xi$ defined as 
a ratio of integrated luminosities of $\bar{\nu}_x$ to $\bar{\nu}_e$;
\begin{equation}
\xi \equiv
\frac {E^{tot}_{\bar{\nu}_{x}}}{E^{tot}_{\bar{\nu}_{e}}},
\end{equation}
where $E^{tot}_{\bar{\nu}_{\alpha}} \equiv \int L_{\bar{\nu}_{\alpha}}dt$
and $\xi=1$ in the equipartition limit. 
It quantifies the departure from equipartition of integrated 
luminosities to $\bar{\nu}_e$ and $\bar{\nu}_x$.
On the other hand, we ignore the latter two effects in this paper.  We
feel that taking into account the uncertainly in $\xi$ gives us a
reasonable framework at least as a first approximation; the other
effects are not very sizable, $\lsim 10$ \%, and may cancel with 
each other, e.g., effects of \cite{horo} and \cite{BJKRR} on 
temperatures of $\nu_{x}$ to $\bar{\nu}_{x}$.
Therefore, we treat 
$E^{tot}_{\bar{\nu}_{e}}$, 
$T_{\bar{\nu}_e}$,
$E^{tot}_{\nu_{x}} = E^{tot}_{\bar{\nu}_{x}}$, and 
$T_{\nu_x} = T_{\bar{\nu}_x}$ 
as free fit parameters.

\section{Analysis method}

Taking the best-fit values of mixing parameters from the latest
analysis of the solar neutrino data~\cite{afterSNO},
we compute the conversion probability for 
$\bar{\nu}_e \leftrightarrow \bar{\nu}_x$ oscillations in 
the SN envelope, following the
prescription in Ref.~\cite{Kachelriess:2001bs}, assuming an
approximate power-law density profile, $\rho(r) \sim r^{-3}$.
For simplicity, we neglect possible Earth matter
effects~\cite{EarthSN} which will depend on the location 
of the detectors.
In the context of three-neutrino flavor mixing, our treatment 
applies to the normal mass ordering ($\Delta m^2_{atm} > 0$) 
and the inverted mass ordering ($\Delta m^2_{atm} < 0$) with
non-adiabatic high-density resonance \cite{MN00,DS99}, where 
$\Delta m^2_{atm}$ denotes the atmospheric neutrino mass squared 
difference. Further implications of our proposal for the case 
of generic three flavor mixing will be given in Ref.~\cite{MNTV}.

We take, for definiteness, the SK and the HK detectors and assume
their fiducial volumes as 32 kton and 1 Mton, respectively. We assume
$\bar{\nu}_e$'s will come from a galactic SN located at 10 kpc from
the Earth.
We only consider $\bar{\nu}_e$ CC reaction on proton and neglect 
contribution from ${\nu}_\alpha e$ elastic scattering 
and CC reactions on oxygen in our analysis. It should give 
a good approximation because these processes have very 
small cross sections.
The effect of weak magnetism is taken into account \cite{vogel}. 
Without knowing the detection efficiency and energy resolution 
expected at HK, we assume that they are the same
as those in SK and set the threshold energy to 5 MeV \cite{MNTV}.

Lacking real galactic SN neutrino data at hand, we generate 
an artificial data set by adopting the model spectra as described 
before. 
We first define arbitrarily a set of initial values for the four 
relevant astrophysical SN parameters 
$ \alpha^0 \equiv \left\{ E_b^0, ~\langle
 E_{\bar{\nu}_e} \rangle^0, ~\tau^0_E, 
~\xi^0 \right\} $ as
``true values'' given by nature.
The data thus generated, $N_i^{\rm{obs}}\equiv N_i(\alpha^0)$, 
are taken as the ``observed'' values. 
To quantify how well a single FD distribution can be 
discriminated from a superposition of two FD distributions
we employ a $\chi^2$ minimization fit in the space of four 
astrophysical parameters 
$ \alpha \equiv \left\{ E_b, ~\langle E_{\bar{\nu}_e} \rangle, 
~\tau_E, ~\xi \right\} $. 
The $\chi^2$ function is defined as 
\begin{equation}
\chi^2 \equiv 2 \times \sum_{i=1}^{N_{bin}} 
\left\{N_i(\alpha)-N_i(\alpha^0)+N_i(\alpha^0)\ln[N_i(\alpha^0)/
N_i(\alpha)]\right\},
\end{equation}
where $N_{bin}=20$.
For simplicity, we present below the results obtained when
event-by-event fluctuations of the artificial data set are 
neglected. However, we have explicitly verified that our results 
agree well with the ones obtained by generating 10,000 sets of 
Gaussian fluctuated data. 

Current solar neutrino data strongly favor the large mixing angle
(LMA) MSW solution. 
Therefore, we focus  on the LMA solution.  It also gives a conservative
estimate of the oscillation effect compared to the LOW and the VAC
solutions because the effect is larger for larger mixing angles.

\section{Results and discussions}

\begin{figure}[h]
\vspace*{2cm}
\includegraphics{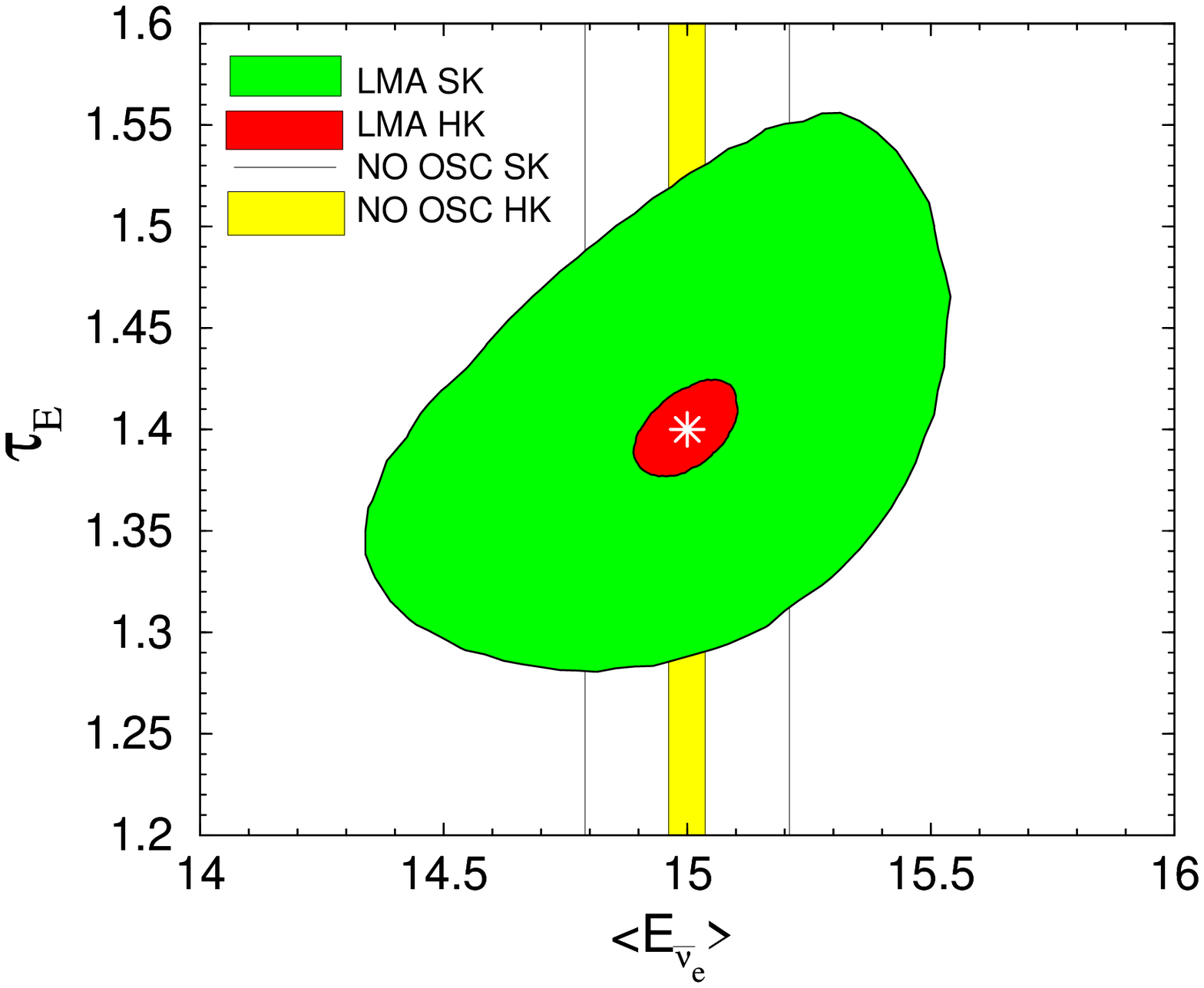}
\includegraphics{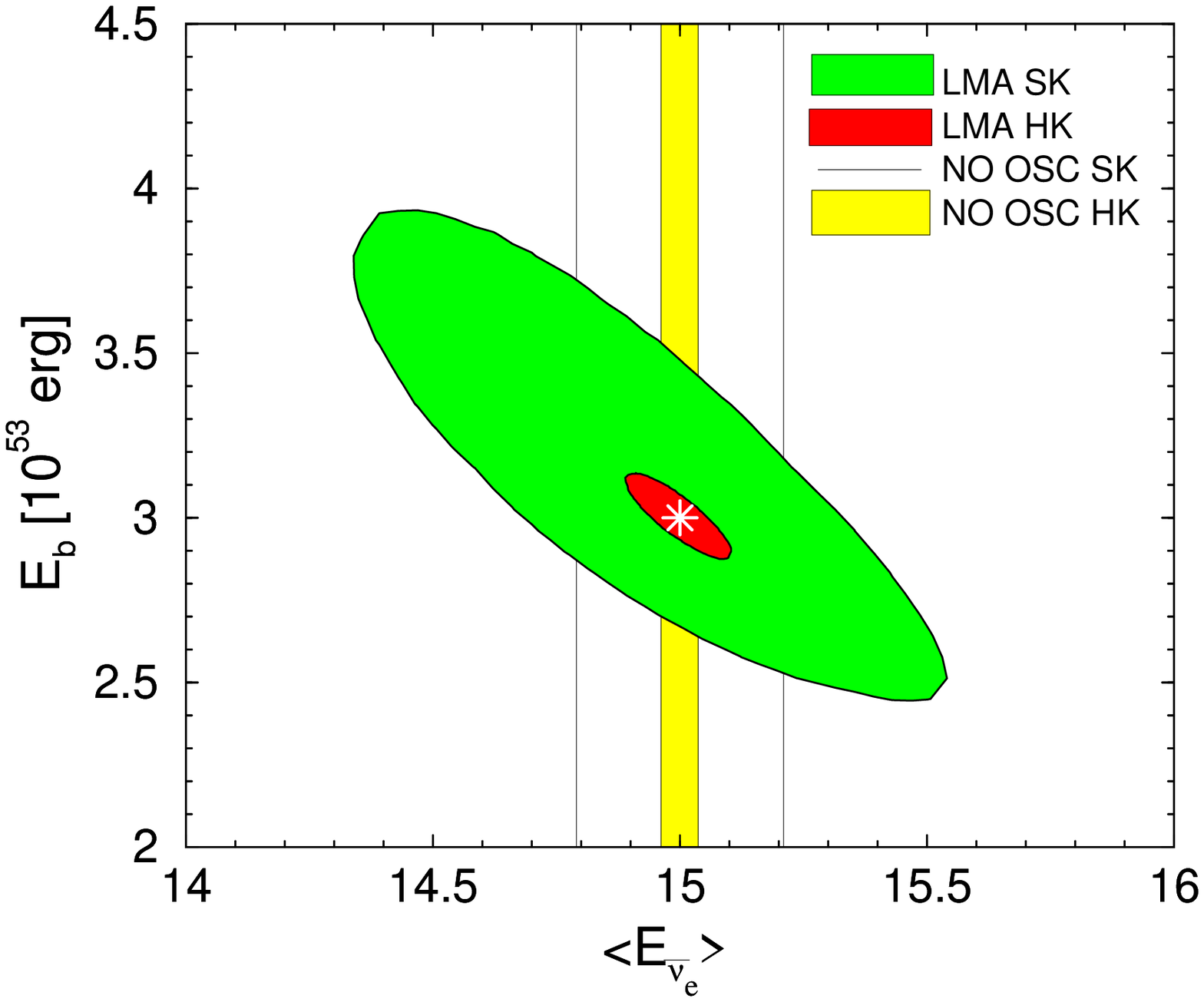}
\vspace{6.0cm}
\caption[]{
  Extracting the astrophysical parameters. The figure shows 3 $\sigma$
  contours assuming $\langle E_{\bar{\nu}_e} \rangle^0 =15$ MeV,
  $\tau_E^0$ = 1.4, $E_b^0 = 3 \times 10^{53}$ erg, and $\xi^0$ = 0.5
  for the SK and HK detectors assuming the LMA solution to the solar
  neutrino problem.  Best fits are indicated by the stars.
}
\label{FvsNF1}
\vglue 0.5cm
\end{figure}

We show in Fig. \ref{FvsNF1} the $3~\sigma$ C.L. allowed parameter
region obtained in the $\langle E_{\bar{\nu}_e} \rangle - \tau_E$ and
$\langle E_{\bar{\nu}_e} \rangle - E_b$ planes for SK and HK
detectors by assuming the LMA solution, taking the best-fit value for
mixing angle, $\tan^2{\theta} =$ 0.42~\cite{afterSNO}.  We also
present results for the no-oscillation case for comparison.  We set
the initial astrophysical parameters as $\langle
E_{\bar{\nu}_e}\rangle^0 = 15$ MeV, $\tau_E^0$ = 1.4, $E_b^0 = 3
\times 10^{53}$ erg, and $\xi^0$ = 0.5.

Fig. \ref{FvsNF1} demonstrates that we can extract the both
$\bar{\nu}_e$ and $\bar{\nu}_x$ temperatures in the presence of large
mixing oscillations.
The accuracies we can achieve with the LMA best-fit parameters are
$\Delta \tau_E/\tau_E \sim $ 9 (1.5) \%, $\Delta \langle
E_{\bar{\nu}_e} \rangle/ \langle E_{\bar{\nu}_e} \rangle \sim $ 4 (1)
In contrast, in the absence of oscillation there is no sensitivity in
$E_b$ due to the inability of determining $\nu_x$ and $\bar{\nu}_x$
luminosities.
The improvement in the accuracy of the determination of these
quantities for the oscillation case is remarkable, especially with HK.

\begin{figure}[h]
\vspace*{3.0cm}
\includegraphics{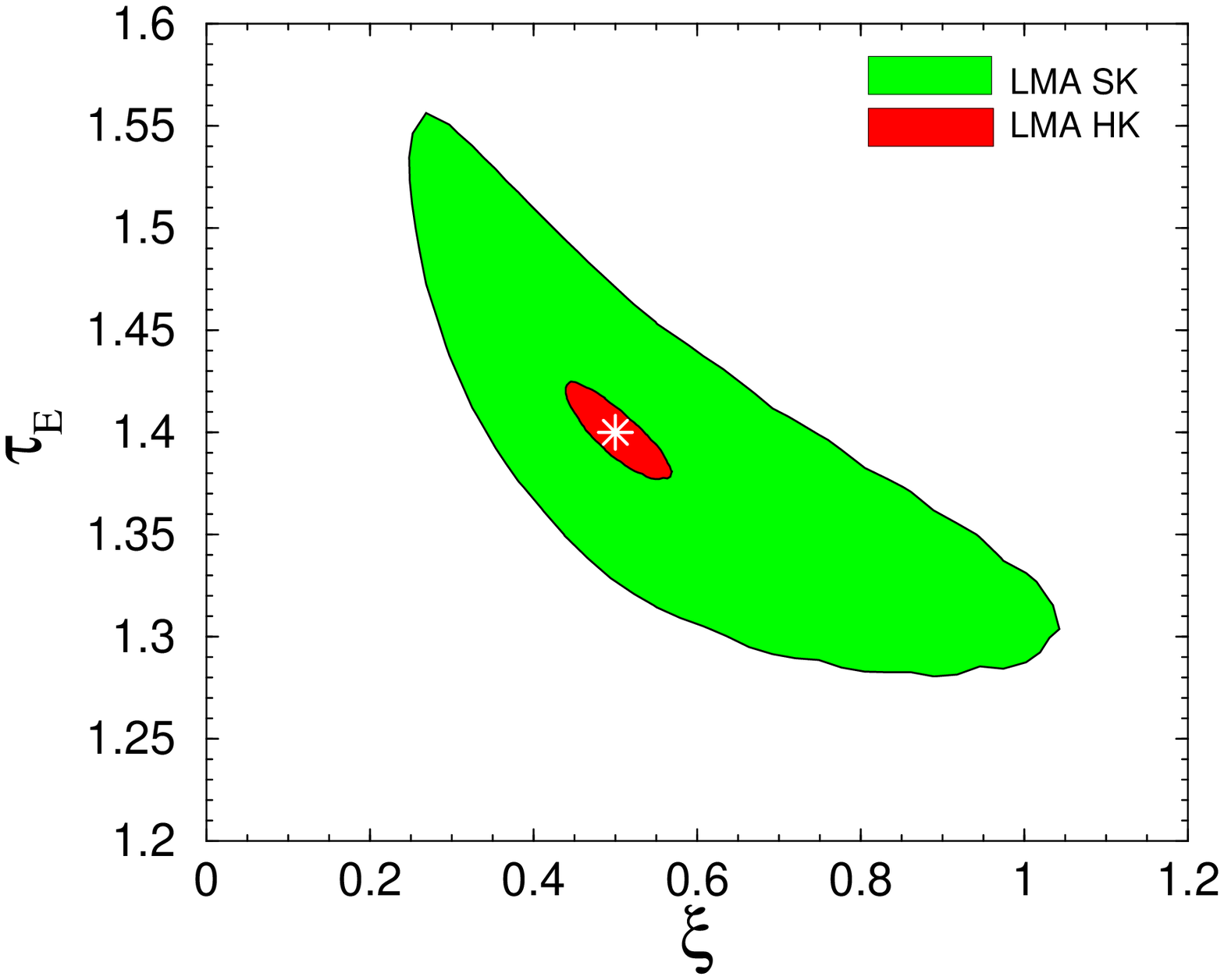}
\includegraphics{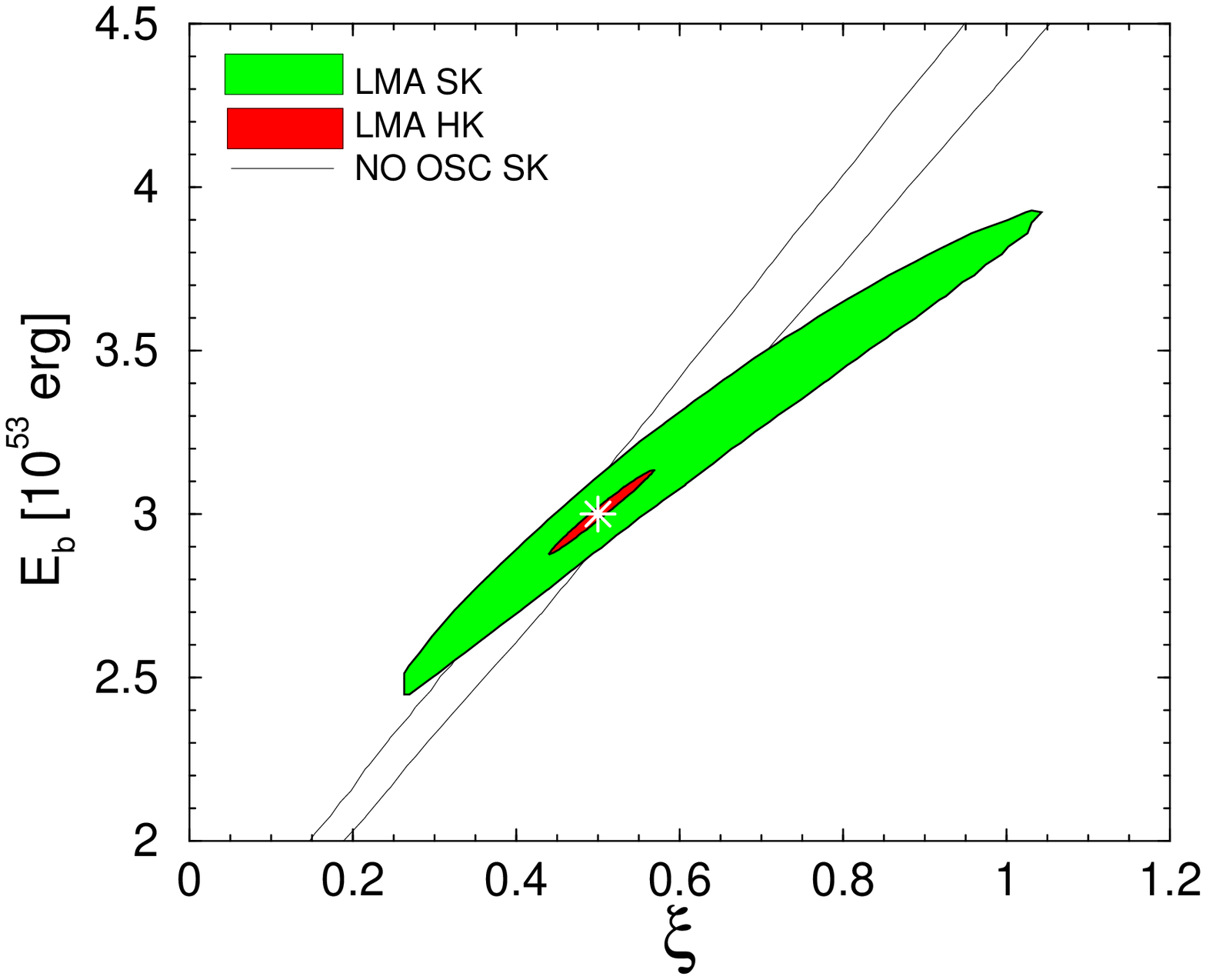}
\vspace{6.0cm}
\caption[]{
Determination of the non-equipartition parameter $\xi$ 
and its correlations with $\tau_E$ and $E_b$.
Same assumptions as in Fig. \ref{FvsNF1}.  
}
\label{FvsNF2}
\vglue 0.5cm
\end{figure}

Let us now examine how accurately we can determine the
equipartition-violation parameter $\xi$ and its correlation with
$\tau_E$ and $E_b$.  This is shown in Fig. \ref{FvsNF2}.
We note that the accuracy of $\tau_E$ determination is remarkably good
in spite of the rather poor accuracy in the knowledge of $\xi$. It
should be noticed that Fig.~\ref{FvsNF2} demonstrates that HK can do
a much better job for $\xi$, $\Delta \xi/\xi \sim $ 10 \%.
We emphasize that without having sensitivity to $\xi$ we can not
determine $E_b$ in a good accuracy, as they are strongly correlated
(See the right panel of Fig.~\ref{FvsNF2}.)
We finally note that HK's enormous sensitivity $\sim 10$ \% may mean
one could potentially examine the problem by treating accreting and
thermal phase separately under the present approximations.

The negative correlation between $\xi$ and $\tau_E$ can be understood
due to the fact that the effect of lowering $\xi$, which implies a
decrease of the relative contribution of $\bar{\nu}_x$, can be
compensated by increasing $\tau_E$ to keep the higher energy tail of
the observed spectrum similar. 
On the other hand, the strong positive correlation between $\xi$ and
$E_b$ just reflects the relationship $E_b = 2 (1 + 2 \xi)
E^{tot}_{\bar{\nu}_{e}}$, which is valid under our approximation
$E^{tot}_{\nu_{e}} = E^{tot}_{\bar{\nu}_{e}}$. The correlation is
robust and exists with and without oscillations as indicated in
Fig.~2.

We have also examined the case of a smaller value of the initial
$\bar{\nu}_x$ energy, $\tau_E^0$ = 1.25 to verify the robustness of
our results.
We found that although the sensitivity to $E_b$ determination worsens
by a factor of $\sim$ 2, we still have a reasonable sensitivity to
$\tau_E$, $\Delta \tau_E/\tau_E \sim $ 14 \% (2.5 \% for HK),
excluding the case of $\tau_E$ = 1 at 3 $\sigma$.
See Ref.~\cite{MNTV} for details.

We have verified that our results do not change very much even if we
relax our assumptions by taking into account deviations from power-law
density profile of progenitor star, or from the precise best-fit
values of solar neutrino mixing parameters we have adopted.  The
inclusion of possible Earth matter effects for a given experiment will
imply some regeneration effect.  Although this will somewhat weaken
our results, the effect is rather small in practice.  Additional
details related to the robustness of our method can be found in
Ref.~\cite{MNTV}.

In summary, we have suggested a simple but powerful way of extracting
separately the temperatures of both $\bar{\nu}_e$ and
$\bar{\nu}_{\mu(\tau)}$ as well as their integrated luminosities by
analyzing $\bar{\nu}_e$ events that would be recorded by massive water
Cherenkov detectors in the event of a galactic supernova explosion.
In particular, an extraordinary power of megaton-class detectors 
(Hyper-Kamiokande or UNO) are noticed with regard to the 
determination of the integrated luminosities of $\bar{\nu}_e$ and
$\bar{\nu}_{\mu(\tau)}$ as independent fit parameters.
We stress that the large mixing between $\nu_e$ and $\nu_{\mu(\tau)}$,
which is clearly indicated by the current solar neutrino data, is
essential in determining temperature as well as luminosity of
$\bar{\nu}_{\mu(\tau)}$, and this must have more profound implications
which await further investigation.

\vskip 0.8cm

\noindent
Note added:

When the first version of this paper was ready for submission to 
the electronic archive we became aware of the paper by 
Barger {\it et al.} \cite {BMW01} who pursued the similar 
strategy as ours. However, they do not treat the violation of 
equipartition of energies as a fit parameter.

\begin{acknowledgments}
  We thank T. Totani and J. Beacom for useful correspondences.  This
  work was supported by the European Commission under contract
  HPRN-CT-2000-00148, by the European Science Foundation
  \emph{Neutrino Astrophysics Network}, by Spanish MCyT grant
  PB98-0693; by the Brazilian FAPESP Foundation, and by the Japan
  Ministry of Education, Culture, Sports, Science and Technology grant
  No. 12047222.  R.T. was supported by Generalitat Valenciana and
  Deutsche Forschungsgemeinschaft.
\end{acknowledgments}


\end{document}